\documentclass[12pt]{pos}

\usepackage{graphicx,epsfig,amsmath,amssymb,amsfonts}
\usepackage{latexsym,cite,eucal}
\newcommand{\beq}{\begin{equation}}
\newcommand{\eeq}{\end{equation}}
\newcommand{\nn}{\nonumber \\}
\newcommand{\fs}{\,.}
\newcommand{\co}{\,,}

\renewcommand{\dag}{^\dagger}

\newcommand{\eps}{\epsilon}

\newcommand{\mr}{\mathrm}

\newcommand{\MeV}{\,\mbox{MeV}}

\newcommand{\mpn}{M_{\pi^0}}
\newcommand{\mpc}{M_\pi}
\newcommand{\pin}{\pi^0}

\newcommand{\po}{p_2}
\newcommand{\pn}{p_1}

\newcommand{\Fnr}{\mathcal{F}}
\newcommand{\mpr}{m_p}
\newcommand{\mn}{m_n}

\newcommand{\vq}{{\bf q}}
\newcommand{\vk}{{\bf k}}
\newcommand{\kt}{k_0}
\newcommand{\knl}{k_1}
\newcommand{\qstar}[1]{h^{#1}(s,m_c,M_{\pi^d})}

\title{Photoproduction of neutral pions}

\ShortTitle{Photoproduction of neutral pions}

\author{\speaker{Andreas Fuhrer}\\
        Department of Physics, University of California at San Diego,
        La Jolla, CA 92093\\
        E-mail: \email{afuhrer@physics.ucsd.edu}}

\abstract{
The pion and nucleon mass differences generate a very
    pronounced cusp in the photoproduction reaction of a single
    neutral pion on the proton. A nonrelativistic effective field theory to
    describe this reaction is constructed. The approach is rigorous in the sense that it is an effective
field theory with a consistent power counting scheme. 
Expressions for the $S$- and $P$-wave multipole amplitudes at one loop
are given. The relation of the phase of the electric multipole
$E_{0+}$ to the phase of the $S$-wave of $\pi^0 p \to \pi^0 p$
scattering is discussed.
}

\FullConference{6th International Workshop on Chiral Dynamics\\
                 July 6-10 2009\\
                 Bern, Switzerland}

\begin{document}

\section{Introduction}
The photoproduction reaction of neutral pions on the proton is a
reaction which shows a strong effect due to isospin breaking. The electric multipole $E_{0+}$ exhibits an
exceptionally strong cusp at the $\pi^+ n$ threshold (see for instance
Ref.~\cite{schmidt}). The strength of this cusp is intimately related to the charge
exchange scattering length of pion-nucleon scattering. Therefore,
accurate experimental data of the photoproduction reaction allows one
to access the pion-nucleon scattering lengths. 

Along the lines of previous work which analyzed the cusps in $K \to 3\pi$
and $\eta' \to \eta \pi \pi$ decays \cite{CGKR,BFGKR,KS}, we construct a nonrelativistic theory which provides
a rigorous framework to describe the structure of the cusp order by
order in a perturbative expansion in terms of small momenta and effective
range parameters of  pion-nucleon scattering as well as threshold
parameters of the photoproduction reaction \cite{fuhrer}. By construction, the
theory correctly reproduces the low-energy singularities in the
Mandelstam plane.

The cusp in neutral pion photoproduction has been studied
before. Ref.~\cite{BKM2} introduces a two-parameter model, which captures
the most important leading effect of the cusp. In
Ref.~\cite{bernstein}, a coupled channel $S$-matrix approach is used
to investigate the cusp structure.

\section{Multipole decomposition}
Some basic relations and definitions used in the
analysis of pion photoproduction are collected.
We calculate the matrix element for the process  $p(p_1)+\gamma(k) \to
p(p_2) + \pi^0(q)$ at leading order in the
electromagnetic coupling $e$, 
\begin{align}
\langle p_2, q\,\, \mathrm{out} | p_1, k\,\, \mathrm{in} \rangle &=  -i (2\pi)^4
\delta^{(4)}(P_f-P_i) \, \bar{u}(\po,t') \eps_\mu J^\mu u(\pn,t) \co
\end{align}
where $P_i$ and $P_f$ denote
the total four momentum in the initial 
and in the final state, respectively and $\eps^\mu$ stands for the
polarization vector of the photon.

To analyze the photoproduction reaction of pions, electric
and magnetic multipoles are usually introduced. To this end, the amplitude
is written in terms of two component spinors $\xi_t$ and Pauli matrices
$\tau^k$ \cite{CGLN},
\begin{align}\label{eq:nrr}
\CMcal{M} &= 8\pi \sqrt{s}\, \xi\dag_{t'}\, \Fnr\,
\xi_{t} \co \nn
\Fnr &= i \boldsymbol\tau \cdot \boldsymbol\epsilon\,  \Fnr_1+
\boldsymbol\tau\cdot \hat{{\bf q}}\, \boldsymbol\tau\cdot (\hat{{\bf k}}\times
\boldsymbol\epsilon)\,\Fnr_2 +i \boldsymbol\tau\cdot \hat{{\bf k}}\,
\hat{{\bf q}}\cdot \boldsymbol\epsilon\, \Fnr_3 +i \boldsymbol\tau
\cdot\hat{{\bf q}}\, \hat{{\bf q}}\cdot \boldsymbol\epsilon\, \Fnr_4
\fs
\end{align}
The hat denotes unit vectors. The $\Fnr_i$ are decomposed into
electric and magnetic multipoles with the help of 
derivatives of the Legendre polynomials $P_l(z)$ \cite{CGLN},
\begin{align}
\Fnr_1 &= \sum_{l=0}\,
    [lM_{l+}+E_{l+}]P_{l+1}'(z)+[(l+1)M_{l-}+E_{l-}]P_{l-1}'(z) \co \nn
\Fnr_2 &= \sum_{l=1}\,[(l+1)M_{l+}+lM_{l-}]P_l'(z) \co \nn
\Fnr_3 &=
\sum_{l=1}\,[E_{l+}-M_{l+}]P_{l+1}''(z)+[E_{l-}+M_{l-}]P_{l-1}''(z)
\co \nn
\Fnr_4 &= \sum_{l=1}\,[M_{l+}-E_{l+}-M_{l-}-E_{l-}]P_l''(z) \fs
\end{align}
The discussion is restrained to the
center of mass frame in the rest of the article.

\section{Nonrelativistic framework}

To describe the behavior of the multipoles close to threshold -- where the energy of the produced
pion and of the proton are small -- a nonrelativistic
calculation is justified. Furthermore, it offers the advantage that
all the masses can be set to their physical value. Therefore, all the
poles and branch points appear at the correct place in the Mandelstam
plane. Moreover, the interaction of the nucleon and the pion is described by effective
range parameters, which allows one to directly access the pion-nucleon scattering
lengths.

The covariant formulation of nonrelativistic field theories introduced in\linebreak
Refs.~\cite{CGKR,BFGKR,KS} is used here since it incorporates the correct
relativistic dispersion law for the particles.  The nonrelativistic proton, neutron and pion fields are denoted by
$\psi$, $\chi$ and $\pi_k$, respectively. The kinetic part of the Lagrangian after minimal substitution takes
the form (see Ref.~\cite{BFGKR2})
\begin{align}
\CMcal{L}_{kin} &= \sum_\pm\Bigl(i\pi_\pm^\dagger D_t{\cal W}_\pm\pi_\pm
-i(D_t{\cal W}_\pm\pi_\pm)^\dagger\pi_\pm-2\pi_\pm^\dagger {\cal W}_\pm^2\pi_\pm\Bigr)\nn
&+ i\psi^\dagger D_t{\cal W}_p\psi
-i(D_t{\cal W}_p\psi)^\dagger\psi-2\psi^\dagger {\cal W}_p^2\psi \nn
&+ 2 \chi\dag W_n (i \partial_t-W_n)\chi + 2 \pi_0\dag W_0 (i\partial_t-W_0)\pi_0 \co
\end{align}
with
\begin{align}
W_0 &= \sqrt{\mpn^2-\triangle} \co &W_n &= \sqrt{m_n^2-\triangle}\co
&D_t\pi_\pm &= (\partial_t\mp ieA_0)\pi_\pm \co\nn
D_t\psi &= (\partial_t- ieA_0)\psi \co &{\cal W}_\pm &= \sqrt{M_\pi^2-{\bf D}^2} \co &{\cal W}_p &=
\sqrt{\mpr^2-{\bf D}^2} \co \nn 
{\bf D}\pi_\pm &= (\nabla\pm ie{\bf A})\pi_\pm \co &
{\bf D}\psi &= (\nabla + ie{\bf A})\psi \fs
\end{align}
Note that since the photon is treated as an external field, its
kinetic term is absent.

\section{Power counting}

Close to threshold, the momenta of the incoming proton and photon are of the 
order of the pion mass whereas the outgoing particles have very
small momenta. Therefore, we count momenta of the outgoing pion and
the outgoing proton as a small quantity of
$O(\eps)$ and the momenta of the incoming proton and of the photon as
$O(1)$. All the masses are counted as $O(1)$. The mass differences
of the charged and neutral pion, $\Delta_\pi \equiv \mpc^2-\mpn^2$ and of
the proton and the neutron, $\Delta_N \equiv \mn^2-\mpr^2$ are counted as $O(\eps^2)$.
At first sight, this counting scheme seems to lead to infinitely many terms already in the leading
order $p +\gamma \to p +\pin$ Lagrangian $\CMcal{L}_\gamma$ because derivatives on the
incoming fields are not suppressed.
However, since the the modulus of the momentum of the incoming
particles, $|{\bf k}|$, can be expanded in the small momentum $|{\bf q}|$,
\begin{align} \label{eq:k}
|{\bf k}| &= \sum_{n}k_n \vq^{2n} \co &\kt &=
\frac{\mpn}{2}\, \frac{2+y}{1+y} \co &\knl &= \frac{y^2+2y+2}{4 \mpn (1+y)} \co &y &=
\frac{\mpn}{\mpr} \co
\end{align}
one obtains a valid power counting scheme. The Feynman rule in
momentum space of every operator of order $\epsilon^0$ with a given
arbitrary number of derivatives can be expanded in powers of the small
momentum $\vq$, yielding one term of order $\epsilon^0$ without any
momenta of the incoming fields present and subsequent higher order
terms. Doing this for every operator of $O(\eps^0)$, all the resulting
leading order terms without any momentum dependence can be described
by one operator of order $O(\eps^0)$ in the interaction Lagrangian. The
same procedure leads to finite numbers of operators at any given higher order
in $\eps$. The derivatives on the incoming fields are only
needed to generate unit vectors in the direction of the incoming
photon. This shows that the nonrelativistic theory is not capable of predicting
the dependence on $|{\bf k}|$ even at threshold. \newline
An additional generic parameter $a$ is introduced to count the
pion-nucleon scattering vertices. Every pion-nucleon interaction
vertex counts as a quantity
of order $O(a)$ since the coupling constants are proportional to the
pion-nucleon scattering threshold parameters, which are
small.
The perturbative expansion is therefore a combined
expansion in $\eps$ and $a$.

\section{Interaction Lagrangian}

The Lagrangian needed for the calculation of the amplitudes for pion 
photoproduction reads $\CMcal{L} = \CMcal{L}_{kin}+\CMcal{L}_\gamma+\CMcal{L}_{\pi N}$,
where $\CMcal{L}_{kin}$ denotes the kinetic part, $\CMcal{L}_\gamma$
incorporates the interaction with the photon field and and
$\CMcal{L}_{\pi N}$ describes the
pion-nucleon sector.

In the pion nucleon sector, the leading terms of the Lagrangian have
been given before in Ref.~\cite{LR}. First, some notation is introduced in order to
write the Lagrangian in a compact form. 
For every channel $n$, we collect the charges of the outgoing and the
incoming pions in the variables $v$ and $w$, $(n;v,w)$: $(0;0,0) ,
(1;0,+),\, (2;+,+),\, (3;0,0),\, (4;-,0),\, (5;-,-)$, thereby assigning \linebreak
unique values to the variables $v$ and $w$ once $n$ is given. The
Lagrangian reads
\begin{align}\label{eq:Lpn}
\CMcal{L}_{\pi N} &= \left( \psi\dag \,\, \chi\dag \right)
\left( \begin{array}{cc} T_{\{0,5 \}} & T_{\{1,4\}}
  \\ T_{\{1,4\}}\dag & T_{\{2,3 \}} \end{array}
\right) \left( \begin{array}{c} \psi \\ \chi \end{array} \right) \co
\nn
T_\CMcal{C} &= \sum_{n \in\, \CMcal{C}}\left[ C_n \pi_v\dag \pi_w + D_n^{(1)}\nabla^k
\pi\dag_v \nabla^k \pi_w + D_n^{(2)} \pi\dag_v
\overleftrightarrow{\triangle} \pi_w + i D_n^{(3)} \tau^k
\epsilon^{ijk}\nabla^i \pi\dag_v \nabla^j \pi_w \right] 
\end{align}
with the abbreviation $f\overleftrightarrow{\triangle}g \equiv f \triangle
g + (\triangle f) g$.\newline
For $\CMcal{L}_\gamma$, the photon is treated as an external vector field ${\bf A}$ which is
odd under parity and time-reversal transformations. One obtains for
the gauge invariant Lagrangian
\begin{align}
\CMcal{L}^{(0)}_\gamma &= -i G_0^{(1)}\psi\dag \tau^k \psi\, E^k\,
\pi_0\dag \co \nn
\CMcal{L}^{(1)}_\gamma &= -i G_1^{(2)}\, \psi\dag \tau^k
\psi\, \nabla^j E^k\, \nabla^j \pi_0\dag  + i G_2^{(1)}\, \psi\dag \tau^m \tau^l
\psi\,B^l\, \nabla^m \pi_0\dag \nn
 &-i G_3^{(2)}\, \psi\dag \tau^j \psi\,
\nabla^j E^k\, \nabla^k \pi_0\dag \co \nn
\CMcal{L}^{(2)}_\gamma &= -iG_4^{(3)} \psi\dag \tau^k \psi \nabla^{jl} E^k
\nabla^{jl} \pi_0\dag -i G_5^{(1)} \psi\dag \tau^k \psi E^k \triangle \pi_0\dag \nn
&+i G_6^{(2)} \psi\dag \tau^m \tau^l \psi \nabla^{n} B^l \nabla^{mn}
\pi_0\dag  -i G_7^{(3)} \psi\dag \tau^j \psi \nabla^{jl} E^k \nabla^{kl}
\pi_0\dag \nn 
&-i G_8^{(1)} \psi\dag \tau^j \psi E^k \nabla^{jk}\pi_0\dag \co\nn
\CMcal{L}^{(3)}_\gamma &= -iG_9^{(2)} \psi\dag \tau^k \psi \nabla^j E^k
\triangle \nabla^j \pi_0\dag -i G_{10}^{(4)} \psi\dag \tau^k \psi
\nabla^{lmn}E^k \nabla^{lmn} \pi_0\dag \nn &+ i G_{11}^{(1)} \psi\dag \tau^m
\tau^l \psi B^l \triangle \nabla^m \pi_0\dag + i G_{12}^{(3)}
\psi\dag \tau^m \tau^l \psi \nabla^{in} B^l \nabla^{min} \pi_0\dag
\nn
&-i G_{13}^{(2)} \psi\dag \tau^j \psi \nabla^j E^k \triangle \nabla^k
\pi_0\dag -i G_{14}^{(4)} \psi\dag \tau^j \psi \nabla^{jlm} E^k \nabla^{klm}
\pi_0\dag \nn
&-i G_{15}^{(2)} \psi\dag \tau^j \psi \nabla^l E^k \nabla^{jkl}
\pi_0\dag  \fs
\end{align}
The upper index on the coupling constants is introduced for later convenience.
Here, the notation $\nabla^{i_1i_2\ldots i_k} \equiv \nabla^{i_1}
\nabla^{i_2} \cdots \nabla^{i_k}$ is used. Since the structure of the
Lagrangian for the other required channel $p \gamma \to n \pi^+$ stays the same, one only has to
replace the coupling constants and the field operators,
$\{\psi\dag,\pi_0\dag,G_i^{(n)}\} \to \{\chi\dag,\pi_+\dag,H_i^{(n)} \}$.
The full interaction Lagrangian $\CMcal{L}_\gamma$ is then given by
adding the $\CMcal{L}^{(i)}_\gamma$ of both channels.

\section{Matching relations}

In the pion-nucleon sector, the coupling constants of the
nonrelativistic Lagrangian, $C_i$ and $D^{(k)}_i$ can be expressed in
terms of pion-nucleon scattering lengths of the $S$-wave and $P$-wave,
$a_{0+}$ and $a_{1\pm}$ and effective range parameters $b_{0+}$, respectively. Adopting the notation of
Ref.~\cite{Hohler},
in the isospin limit, the isospin decomposition of the $\pi N$ scattering amplitudes reads
\begin{align}\label{eq:amps}
T_{p\pi^0\to p\pi^0} &= T_{n\pi^0 \to n\pi^0} = T^+ \co &T_{p \pi^0
  \to n\pi^+} &= T_{n\pi^0 \to p
  \pi^-} = -\sqrt{2}\, T^-
\co \nn T_{n\pi^+ \to n \pi^+} &= T_{p\pi^- \to p\pi^-} = T^++T^- \fs
\end{align}
Defining $\CMcal{N} = 4\pi (\mpr+\mpc)$, one finds
\begin{align}\label{eq:matching}
C_0 &= 2\, \CMcal{N}  a_{0+}^+ \co &C_1 &= 2\sqrt{2}\,\CMcal{N}
a_{0+}^-  \co &C_2 &= 2\, \CMcal{N} (a_{0+}^++a_{0+}^-) \co\nn
C_3 &= C_0 \co &C_4 &= C_1 \co &C_5 &= C_2 \fs
\end{align} 
The matching conditions for the $D_i^{(k)}$ are given in a generic
form only. The isospin index of the threshold parameters can be
inferred from Eq.~(\ref{eq:amps})\footnote{Note that we use the Condon-Shortley phase
convention.}.
\begin{align}
D^{(1)}_i &= 2\CMcal{N} (2 a_{1+}+a_{1-}) \co &D^{(2)}_i &= -
\CMcal{N} \left(\frac{a_{0+}}{2\mpr \mpc}+b_{0+} \right) \co \nn
D^{(3)}_i &= 2\CMcal{N}\ ( a_{1-}-a_{1+})\fs
\end{align}
Here, higher order terms in the threshold parameters have been dropped.
The corrections to these relations which appear due to isospin breaking have to be
calculated within the underlying relativistic theory. For the $C_i$, they can be found
in Refs.~\cite{GILMR,MRR,hkm,hkm2}. Note that the second line in
Eq.~(\ref{eq:matching}) is only true in the isospin limit.

The constants $G^{(n)}_i$ and $H^{(n)}_i$ on the other hand are related to the
threshold parameters of the electric and magnetic multipoles of the
pertinent channel. 
In the isospin limit, the expansion of the real part of the multipole
$X_{l\pm}$ close to threshold is written in the form 
\begin{align}\label{eq:thresholdpar}
\mathrm{Re}X_{l\pm}(s) &= \sum_{k=0}^{\infty} \bar{X}_{l\pm,2k}|\vq|^{l+2k} \co
\end{align}
which defines the threshold parameters $\bar{X}_{l\pm,2k}$. In the
following, the relations of the coupling constants $G^{(n)}_i$ to these threshold
parameters is given at leading order in the pion nucleon threshold parameters. Since the
nonrelativistic theory is not suited for the study of the dependence
of the multipoles on $|\vk|$, in this analysis, all vectors $\vk$ are turned into unit
vectors by the pertinent redefinition of the coupling constants,
\beq\label{eq:redef}
 G_i^{(n)} = \CMcal{N}_0 \kt^{-n} G_i \co \qquad \CMcal{N}_0 = 4\pi (\mpr+\mpn)\fs
\eeq
Note that the higher order corrections due to Eq.~(\ref{eq:k}) have to
be taken care of in the matching relations.
Again, these relations pick up isospin breaking corrections which have to be
evaluated in the underlying relativistic theory.

Only the matching equations for the couplings of the Lagrangians
$\CMcal{L}_\gamma^{(0)}$ and $\CMcal{L}_\gamma^{(1)}$ are indicated
here. The remaining relations can be found in Ref.~\cite{fuhrer}. To ease notation, $\bar{X}_{i\pm} \equiv
\bar{X}_{i\pm,0}$ is used.
\begin{align}
G_0 &= 2\bar{E}_{0+} \co  &G_1 &= 6(\bar{E}_{+1}+\bar{M}_{+1})
\co \nn
G_2 &= -2(\bar{M}_{-1}+2 \bar{M}_{+1}) \co  &G_3 &= 6
(\bar{E}_{1+}-\bar{M}_{1+}) \fs
\end{align} 
For the coupling constants $H_i$ the algebraic form of the relations is identical. However, the multipoles of
the pertinent channels appear and the masses in Eq.~(\ref{eq:redef})
have to be adjusted.

All coupling constants are assumed to be real. See Ref.~\cite{fuhrer}
for a discussion of this issue.

\section{Results}

In the following, we provide the expressions for the electric and
magnetic multipoles $E_{l +}$ for $l=0,1$ and $M_{l \pm }$ for $l =
1$. The result is written in the form 
\begin{align}\label{eq:result}
X_{l,\pm}(s) &= X_{l\pm}^{\mathrm{tree}}(s)+X_{l\pm}^{\mathrm{1
    Loop}}(s) +X_{l\pm}^{\mathrm{2
    Loop}}(s) \cdots
\end{align}
where $s = (p_1+k)^2$ and the ellipsis denote higher order terms in the expansion in
$\eps$ and $a$.

\subsection{Tree-level}

The tree level result can be written in the form $X^{\mathrm{tree}}_{l\pm}(s) = X^t_{l\pm}\vq^l + X^t_{l\pm,2}
\vq^{2+l} + \cdots$ with the coefficients 
\begin{align}\label{eq:tree}
E^t_{0+} &= G_0 \co & 3 E^t_{0+,2} &=
G_4-3 G_5+G_6-G_8 \co \nn
6 M^t_{1+} &= G_1-G_3 \co &M^t_{1+,2} &=
-\tfrac{1}{6}G_9+\tfrac{1}{10}G_{10}+\tfrac{1}{15}G_{12}+\tfrac{1}{6}G_{13}-\tfrac{1}{30}G_{14} \co
\nn
 3 M^t_{1-} &= G_3-G_1-3G_2 \co & M^t_{1-,2} &=
 \tfrac{1}{3}G_9-\tfrac{1}{5}G_{10}+G_{11}-\tfrac{1}{3}G_{12} \nn
&&& -\tfrac{1}{3}G_{13}+\tfrac{1}{15}G_{14} \co
 \nn
 6 E^t_{1+} &= G_1+G_3 \co &E^t_{1+,2} &=
 -\tfrac{1}{6}G_9+\tfrac{1}{10}G_{10}+\tfrac{1}{15}G_{12}-\tfrac{1}{6}G_{13}\nn
&&& +\tfrac{1}{30}G_{14}-\tfrac{1}{15}G_{15} \fs
\end{align}
One observes that $D$-waves appear naturally at order $\epsilon^2$ in
this framework (see also Ref.\cite{FBD}).

\subsection{One-loop}

All the one-loop contributions are proportional to the basic integral
\begin{align}
J_{ab}(P^2) &= \int \frac{d^Dl}{i(2\pi)^D} \frac{1}{2 \omega_a({\bf
    l})2\omega_b({\bf P}-{\bf l}) }\, \frac{1}{( \omega_a({\bf
    l})-l_0 ) (\omega_b({\bf P}-{\bf l}) -P_0 +l_0 ) } \co \nonumber
\end{align}
\begin{align}
\omega_\pm({\bf  p}) &= \sqrt{\mpc^2+ {\bf p}^2} \co  &\omega_i({\bf p}) &=
\sqrt{m_i^2+ {\bf p}^2}\co \qquad i=n,p  \nn
\omega_0({\bf p})  &= \sqrt{\mpn^2+ {\bf p}^2}\co  &P^2 &= P_0^2-{\bf
  P}^2 \fs
\end{align}
In the limit $D \to 4$,
\beq\label{eq:loopfunc}
J_{ab}(P^2) = \frac{i}{16\pi
  s}\sqrt{(s-(m_a+M_{\pi^b})^2)(s-(m_a-M_{\pi^b})^2)}\co 
\eeq
which is a quantity of order $\eps$.
\begin{figure}
\centering
\begin{tabular}{cc}
\includegraphics[height=1.4cm]{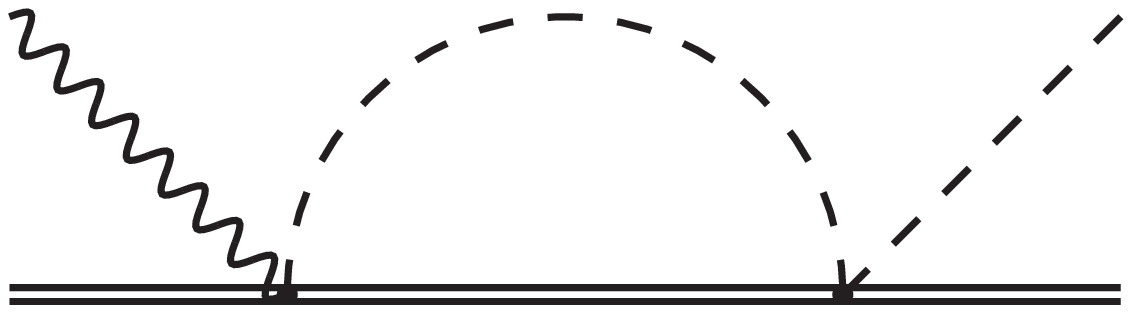}&\includegraphics[height=1.4cm]{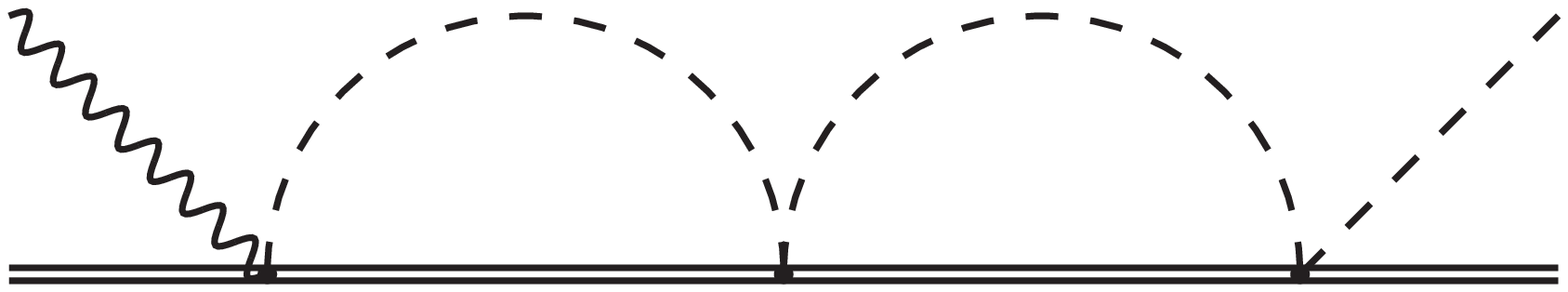}
\end{tabular}
\caption{One- and two loop topologies needed to calculate the
  amplitude. The double line
  generically denotes a nucleon, the dashed line a pion and the
  wiggly line indicates the external electromagnetic field.}\label{fig:diags}
\end{figure}
The one-loop result up to and including order $O(a \epsilon^4)$ reads
\begin{align} \label{eq:1loop}
\left( \begin{array}{c} E_{0+}^{\mathrm{1 Loop}}(s) \\ \frac{1}{|\vq|}M_{1+}^{\mathrm{1
      Loop}}(s) \\ \frac{1}{|\vq|} M_{1-}^{\mathrm{1 Loop}}(s)
  \\ \frac{1}{|\vq|} E_{1+}^{\mathrm{1 Loop}}(s) \end{array}
  \right) &= \left( \begin{array}{cc}  P_{11} & P_{12}\\ P_{21}  & P_{22} \\ P_{31} &
     P_{32} \\ P_{41} &
     P_{42}  \end{array}
  \right) \left( \begin{array}{c}  J_{p0}(s) \\ J_{n+}(s)
     \end{array} \right) \fs
\end{align}
The elements $P_{ik}$ are functions of the pion momentum $\vq$
and the coupling constants of the
Lagrangian,
\begin{align}\label{eq:1loopcoef}
P_{11} &= G_0 C_0 + \vq^2 \left(C_0 E^{(p0),t}_{0+,2}-2 D_0^{(2)}
  G_0\right) \co \nn  P_{12} &= C_1 H_0 + \qstar{2} \left(C_1 E^{(n+),t}_{0+,2}-D_1^{(2)} H_0 \right)-\vq^2 D_1^{(2)} H_0 \co\nn
18 P_{21} &= \vq^2 \left(D^{(1)}_0-D^{(3)}_0\right)( G_1-G_3) \co\nn
18 P_{22} &=  \qstar{2} \left(D^{(1)}_1-D^{(3)}_1 \right) (
H_1-H_3) \co \nn
9 P_{31} &= \vq^2 \left(D^{(1)}_0+2D^{(3)}_0 \right)(G_3-G_1-3G_2) \co\nn
9 P_{32} &= \qstar{2} \left(D^{(1)}_1+2D^{(3)}_1\right)(H_3-H_1-3H_2)\co \nn
18 P_{41} &= \vq^2 \left(D^{(1)}_0-D^{(3)}_0\right)(G_1+G_3) \co\nn
18 P_{42} &= \qstar{2} \left(D^{(1)}_1-D^{(3)}_1\right)(H_1+H_3) \co
\end{align}
where $E^{(c),t}_{0+,2}$ denotes the pertinent coefficient of the tree
level result of channel $(c)$, see Eq.~(\ref{eq:tree}), and $\qstar{2}$ is given by
\begin{align}
\qstar{2} &= \frac{\left(s-(m_n+M_{\pi^+})^2\right)\left(s-(m_n-M_{\pi^+})^2\right)}{4s} \co
\end{align}
which is a quantity of order $\epsilon^2$. Eq.~(\ref{eq:1loop}) and
(\ref{eq:1loopcoef}) clearly show the advantage of the nonrelativistic
description: The strength of the cusp at leading order is parameterized in terms of the
coupling constant $C_1$ and the ratio $H_0/G_0$. 

\section{Phase of $\pi^0 p \to \pi^0 p$ scattering}

In an isospin symmetric world, the phase of the multipole $E_{0+}$ is
directly related to the phase shift of the $S$-wave of pion-nucleon scattering
by virtue of the Fermi-Watson theorem \cite{FW}. In
Ref.~\cite{bernstein}, it is shown with a coupled channel $S$-matrix
approach that to leading order in $e$, below the $\pi^+ n$
 threshold, the phase of the $S$ wave of $\pi^0 p \to \pi^0 p$
 scattering is equal to the phase of $E_{0+}$,
\begin{align}
\tan \delta_{p \pi^0 \to p \pi^0} &= \tan
\frac{\mathrm{Im\,}E_{0+}}{\mathrm{Re\,}E_{0+}} \equiv \tan \delta_{E_{0+}}  \fs
\end{align}
The framework developed here allows one to test this statement order
by order in the perturbative expansion. To this end, the phase of $E_{0+}$ below the second
threshold is calculated up to and including $O(a^2 \eps^4)$,
\begin{align}\label{eq:phase}
\tan\delta_{E_{0+}} &= C_0\, \mr{Im}J_{p0} + C_1^2 J_{n+} \mr{Im} J_{p0} - 2D_0^{(2)}
\mr{Im}J_{p0} \vq^2 \\& -2C_1 D_1^{(2)}J_{n+} \mr{Im} J_{p0} \, \vq^2 - 2C_1 D_1^{(2)}
J_{n+} \mr{Im}J_{p0} \, h^2(s,\mn,\mpc) + \cdots \fs \nonumber
\end{align}
Calculating $\pi^0 p \to \pi^0 p$ scattering to the same order with
the Lagrangian given in Eq.~(\ref{eq:Lpn}), one
finds that the phase of the $S$-wave below the second threshold is indeed equal to
Eq.~(\ref{eq:phase}). However, the main object of interest is the
phase of the $S$-wave of $\pi^0 p \to \pi^0 p$
scattering in the {\it isospin symmetry limit},
\begin{align}
\tan \bar{\delta}_{p\pi^0 \to p\pi^0} &= C_0 \mr{Im}J_{p0} + \frac{C_1^2}{C_0}
\mr{Im}J_{n+} \simeq a_{0+}^+ \vq +2\frac{a_{0+}^{-2}}{a_{0+}^+}\vq + \cdots \co
\end{align}
which does not agree with
$\delta_{p\pi^0\to p\pi^0}$ in the presence of isospin violations
already at leading order.

\section{Summary and conclusion}

  We study the photoproduction reaction of pions on the 
  nucleon using a nonrelativistic framework. The electric
  and magnetic multipoles $E_{l+}$ for $l = 0,1$ and $M_{1\pm}$ are
  calculated in a systematic double expansion in the final state pion-
  and nucleon momenta (counted as a small quantity of order $\epsilon$) and
  the threshold parameters of $\pi N$ scattering (denoted by
  $a$). Explicit representations for the multipoles up to and
  including $\epsilon^3$ and $\epsilon^4 a$ are
  provided. The corresponding two-loop results as well as a
  expressions for the multipole amplitudes in the remaining three
  reaction channels can be found in Ref.~\cite{fuhrer}. 

  The representation is valid in the low energy region,
  at least up to a photon energy in the lab frame of $E_\gamma = 165
  \MeV$. It accurately describes the cusp structure and allows one to
  determine the pion-nucleon threshold parameters from experimental data.

  The relation of the phase of the electric multipole $E_{0+}$ in the
  $(p0)$ channel to the phase of the $S$-wave of $\pi^0 p \to \pi^0 p$
  scattering is discussed in the presence of isospin violation. A
  relation found in earlier work \cite{bernstein} is
  confirmed. We stress that the relation does not allow one to obtain
  the phase of $\pi^0 p \to \pi^0 p$ scattering in the isospin limit.

{\it Acknowledgements.} I would like to thank the organizers for a very
interesting workshop. I am also indebted to J.~Gasser, B.~Kubis, A.~Manohar and
U.-G.~Mei\ss{}ner for informative discussions and B.~Kubis for
comments on the manuscript. This work was supported
in part by the Department
of Energy under Grant 
DE-FG03-97ER40546 and by the Swiss National Science Foundation.

\end{document}